\documentclass[12pt]{article}

\begin{document}
\title{\bf Noncommutative Solitons and the $W_{1 + \infty}$ \\
           Algebras in Quantum Hall Theory}
\author{Chuan-Tsung Chan\thanks{email address: ctchan@ep.nctu.edu.tw} \
         and Jen-Chi Lee\thanks{email address: jcclee@cc.nctu.edu.tw}
\vspace{0.8cm} \\
{\small \it Department of Electrophysics, National Chiao-Tung University}
 \vspace{0.2cm}\\
 {\small \it Hsin-Chu, 300, Taiwan}}
\date{}
 \maketitle

\setlength{\baselineskip}{4ex}
\begin{abstract}
  We show that $U(\infty)$ symmetry transformations of the noncommutative
field theory in the Moyal space are generated by a combination of two
$W_{1+\infty}$ algebras in the Landau problem.
  Geometrical meaning of this infinite symmetry is illustrated by
examining the transformations of an invariant subgroup on the
noncommutative solitons, which generate deformations and boosts of
solitons. In particular, we find that boosts of solitons, which
was first introduced in Ref[22], are valid {\it only} in the
infinite $\theta$ limit.
\vspace{0.4cm}
\end{abstract}

\newpage

\section{Introduction}

   Noncommutative field theory has been an intense topic of research
among string/field
 theory community for the last few years.
   One reason is that such a theory naturally appears in the low-energy
description of
 D-brane dynamics in the presence of a constant Neveu-Schwarz B field
background \cite{Connes:1998cr} \cite{Douglas:1998fm}
\cite{Chu:1999qz} \cite{Chu:2000gi} \cite{Schomerus:1999ug}.
   More recently, soliton solutions of noncommutative scalar
\cite{Gopakumar:2000zd} and
 scalar-gauge \cite{Jatkar:2000ei} \cite{Bak:2000ac} \cite{Harvey:2000jb}
theories were constructed.
   These topological objects were then applied to the construction of
D-branes as solitons of the tachyon field in noncommutative open
string theory
 \cite{Sen:1998sm} \cite{Harvey:2000tv} \cite{Harvey:2000jt}.

   The notion of noncommutative coordinates is not a new idea in physics
community.
   In fact, it has been known for a long time in condense matter physics
that two-dimensional
 quantum Hall system realizes the simplest non-commutative geometry
\cite{Fradkin:1991nr},
 whose appearance can be traced back to the quantum realization of
projective representation
 for the magnetic translation group.
   Consequently, it is not surprising that quantum Hall system offers
some kind of toy
 model in the study of noncommutative string/field theories.
   Recent development of exact realization of the first quantized
Laughlin wave function
 by an abelian noncommutative Chern-Simons theory was along this line of
thought
 \cite{Susskind:2001fb}.

   In this paper, we shall unfold another interesting connection between
noncommutative
 field theory and the quantum Hall theory.
   Namely, we show that $U(\infty)$ symmetry transformations of the
non-commutative field
 theory in the Moyal space are generated by a combination of two
$W_{1+\infty}$ algebras in the Landau problem
\cite{Cappelli:1993yv} \cite{Cappelli:1993ea}
\cite{Cappelli:1993kf} \cite{Iso:1992aa} \cite{Karabali:1994gn}
\cite{Karabali:1994sg}.

 As an application, one can use this algebra as a soliton
 generating algebra to generate non-radially symmetric solitons.
   To get some geometrical meaning of our new solitons, we discuss
   and explicitly calculate an invariant subalgebra of this
   soliton generating algebra and, in particular, use it to
   generate deformations and boosts of solitons.
We find that boosts of solitons, which was first
introduced in Ref[22], are valid {\it only} in the infinite $\theta$
limit. The argument of eqs (12)
and (13) used in Ref[22] was misleading for the finite $\theta$ case.
This can be seen from our calculation that area-preserving, or
$W_{1+\infty}$ symmetry, is violated for the finite $\theta$ case.

  \section{$U(\infty)$ Symmetry in Noncommutative Field theory}

     The two-dimensional noncommutative plane is defined by a commutation
rule between
   its coordinates, $\vec {\bf x} = (\hat x, \hat y)$,
   \begin{equation}
     [\hat x, \hat y] = i \theta.
   \end{equation}

     The operator algebra $\cal A$ defined on this noncommutative plane
is generated by
   the formal power series in the noncommutative coordinates,
   \begin{equation}
     {\cal A} \equiv \{ \hat A ( {\vec {\bf x}} ) | \hat A
     ({\vec {\bf x}})  =
      \sum_{mn} a_{mn} {\hat x}^m {\hat y}^n \}.
   \end{equation}
     The Weyl-ordered operator on the noncommutative plane can be
generated from a function
   defined on the Moyal space,
   \begin{equation}
     \hat \phi (\vec{\bf x}) = \int d^2 \vec x \int
     \frac{d^2 \vec k}{(2\pi)^2}
      e^{i \vec k (\vec{\bf x} - \vec x)}
      \phi( \vec x ),
   \end{equation}
   and we have the following isomorphism between operator product and
star-product,
   \begin{equation}
        ( \hat \phi \cdot  \hat \psi )(   \vec {\bf x} )
       =  \int  d^2 \vec x \int \frac{d^2 \vec k}{(2 \pi)^2}
         e^{i  \vec k (\vec {\bf x}- \vec x)}
             ( \phi * \psi )( \vec x ),
   \end{equation}
   where the star-product is defined as
   \begin{equation}
      \phi (\vec x) * \psi (\vec x) \equiv
        exp\left[ i \frac{\theta \epsilon_{ij}}{2}
\frac{\partial}{\partial y_i}
\frac{\partial}{\partial z_j}
\right]
      \phi(\vec y) \psi(\vec z) \left|_{\vec y = \vec z = \vec x},
\right.
      \hspace{0.5cm} i,j = 1,2.
   \end{equation}

     It is convenient to introduce the annihilation and
   creation operators,
     \begin{equation}
       c           \equiv \frac{\hat x + i \hat y}{\sqrt{2 \theta}},
       \hspace{1cm}
       c^{\dagger} \equiv \frac{\hat x - i \hat y}{\sqrt{2 \theta}},
       \hspace{1.2cm}
       [c, c^\dagger] = 1.
     \end{equation}
     Any operator in $\cal A$  can be expanded in terms of the
occupation-number operator basis,
     \begin{equation}
       \hat A (\vec {\bf x}) =
       \sum_{m,n=0}^\infty a_{mn} \hat \phi_{mn} ,
     \end{equation}
   where the operator basis $\hat \phi_{mn}$ is defined as
     \begin{equation}
       \hat \phi_{mn} \equiv \left| m > < n \right| =
       \frac{(c^\dagger)^m}{\sqrt{m!}} \left| 0 > < 0 \right|
\frac{c^n}{\sqrt{n!}}.
     \end{equation}

      We also take the following definitions of commutative variables,
     \begin{equation}
            z   \equiv \sqrt{\frac{2}{\theta}} ( x + i y ),
\hspace{0.7cm}
       \bar z   =      \sqrt{\frac{2}{\theta}} ( x - i y );
\hspace{1cm}
            k_c \equiv \sqrt{\frac{\theta}{2}} (k_x + i k_y),
\hspace{0.7cm}
       \bar k_c =      \sqrt{\frac{\theta}{2}} (k_x - i k_y).
     \end{equation}

     The action of the noncommutative scalar-gauge theory can be written
as

   \begin{equation}
     L = Tr[ - \frac{1}{4} {\hat F}_{ij} {\hat F}^{ij} + (D_i {\hat
\phi})^\dagger
                                                     (D^i {\hat \phi}) +
V
                                                     (\hat \phi)],
   \end{equation}
   where the field strength tensor is
   \begin{equation}
 F_{ij} \equiv  \frac{\epsilon_{k i}}{i \theta} [\hat x_k, \hat A_j] -
                \frac{\epsilon_{k j}}{i \theta} [\hat x_k, \hat A_i]
   + i g [\hat A_i, \hat A_j]
    \end{equation}
    and the covariant derivatives are
\begin{eqnarray}
  D_i \hat \phi  &\equiv& \frac{\epsilon_{k i}}{i \theta} [\hat x_k, \hat
\phi]-
   i g [\hat A_i , \hat \phi], \\
  D_i \hat \phi  &\equiv& \frac{\epsilon_{k i}}{i \theta}
   [\hat x_k, \hat \phi] - i g \hat A_i  \hat \phi,
   \end{eqnarray}
  for real and complex scalar fields, respectively. In the complex case
eq.(13), for our purpose, we only introduce one piece of gauge field (see
eq.(16)).

    The action in eq.(10) obviously contains the $U(\infty)$ symmetry
which
   will be discussed in the next section. One interesting discovery
   recently was that the theory possesses various soliton solutions
    which
   turns out to be important in many applications \cite{Harvey:2001yn}.
    The original GMS solitons, for example, are radially symmetric
solutions of real scalar
    theory in the infinite $\theta$ limit,
  \begin{equation}
   \phi_n (z, \bar z) = 2 (-1)^n L_n (|z|^2) \exp^{- \frac{|z|^2}{2}},
\hspace{0.7cm} n=0,1,...
  \end{equation}
   which can be constructed from the corresponding projection
   operators in the operator algebra $\cal A$.
     The $U(\infty)$ symmetry mentioned
   above can then be used to generate more general
   non-radially symmetric solitons through the following transformation
   \begin{equation}
    \hat \phi \rightarrow  U \hat \phi U^\dagger.
   \end{equation}

   Similar technique and its generalization \cite{Harvey:2000jb} can be
 applied to
   the case of complex scalar-gauge theory. The transformation for the
 scalar is
   \begin{equation}
    \hat \phi \rightarrow      \hat \phi U.
   \end{equation}
 In this case, for simplicity, we consider only right-handed $U(1)$ gauge
symmetry or half of the $U(\infty)$ symmetry. In the next section, we
will
give an explicit form of these
   transformations on the corresponding Moyal function space.

   \section{$W_{1 + \infty}$ Algebras in Quantum Hall Theory from
$U(\infty)$ Symmetry}

   To calculate eq.(15) in the corresponding Moyal function space,
   we first choose the occupation-number operators as a basis of the
operator algebra
   $\cal A$,
    \begin{equation}
     \hat \phi_{mn} = |m ><n| =
     \frac{ (c^\dagger)^m }{\sqrt {m!}} :\exp[- c^\dagger c]:
     \frac{ c^n }{\sqrt {n!}},
    \end{equation}
   where the normal-ordering has been introduced in the second
   equality. On the other hand, the corresponding function basis
   $\varphi_{mn}$ on the Moyal space can be defined through eq.(3)
   \begin{equation}
     {\hat \phi}_{mn} \equiv \left| m > < n \right| =
      \int d^2 \vec x \int \frac{d^2 \vec k}{(2 \pi)^2}
     e^{i \vec k (\vec {\bf x} - \vec x)} \varphi_{mn} (\vec x).
   \end{equation}

    The momentum space function basis can be calculated to be
 \begin{equation}
  \tilde \varphi_{mn}(\vec k) =
                  \frac{1}{\sqrt{m! n!}} e^{\frac{k_c k_{\bar c}}{2}}
                     \left( i \frac{\partial}{\partial k_c} \right)^m
                     \left( i \frac{\partial}{\partial k_{\bar c}}
\right)^n
                  \tilde \chi_{00}(\vec k),
 \end{equation}
   where
 \begin{equation}
  \tilde \chi_{00}(\vec k) \equiv (2 \pi \theta) e^{-\frac{\theta {\vec
k}^2}{2}}.
 \end{equation}
 Taking Fourier transform and after a series of integration by parts, we
get
 \begin{eqnarray}
 \varphi_{mn}(z, \bar z)&=& \frac{1}{\sqrt{m! n!}}
                         \int \frac{d^2 \vec k}{(2 \pi)^2}
                         \left[ \left( - i \frac{\partial}{\partial k_c}
\right)^m
                         e^{ \frac{1}{2}(k_c k_{\bar c} + i k_c \bar z +
i k_{\bar c}
                         z)} \right]
                         \left( i \frac{\partial}{\partial k_{\bar c}}
\right)^n
                         {\tilde \chi}_{00}(\vec k) \nonumber \\
                     &=& \frac{1}{\sqrt{m! n!}}
                         \left[
                         \left(- \frac{\partial}{\partial z} +
                                 \frac{\bar z}{2} \right) \right]^m
                         \int \frac{d^2 \vec k}{(2 \pi)^2}
                         e^{\frac{1}{2} (k_c k_{\bar c} + i k_c \bar z +
i k_{\bar c}
                         z)}
                         \left( i \frac{\partial}{\partial k_{\bar c}}
\right)^n
                         {\tilde \chi}_{00}(\vec k). \nonumber
 \end{eqnarray}
 Similar procedure can be used to simplify the $k_{\bar c}$ partial
derivatives, and we have
 \begin{eqnarray}
 \varphi_{mn}(z, \bar z) &=& \frac{1}{\sqrt{m! n!}}
                         \left[
                         \left( - \frac{\partial}{\partial z} +
                                  \frac{\bar z}{2} \right) \right]^m
                         \left[
                         \left( - \frac{\partial}{\partial {\bar z}}+
                                  \frac{z}{2} \right)  \right]^n
                         \varphi_{00}(z, \bar z), \nonumber \\
                     &=& \frac{\sqrt{2 \pi \theta}}{\sqrt{m! n!}}
                         (a^{\dagger})^m (b^{\dagger})^n \Phi_{00}(z,\bar
z)
                     = \sqrt{2 \pi \theta} \Phi_{mn}(z, \bar z) .
 \end{eqnarray}
 where $a^\dagger, b^\dagger$ and $\Phi_{00}$ are given by
 \begin{equation}
  a^\dagger \equiv - \frac{\partial}{\partial z} +       \frac{\bar
z}{2}, \hspace{1cm}
  b^\dagger \equiv - \frac{\partial}{\partial {\bar z}}+ \frac{z}{2},
 \end{equation}
 \begin{equation}
     \Phi_{00} (\vec x)
               = \sqrt{\frac{2}{\pi \theta}} \exp [- \frac{{\vec
x}^2}{\theta}].
 \end{equation}
  It is important that $a^\dagger$ and $b^\dagger$ are commuting
 operators, so that their ordering is immaterial.
   The physical meaning of these operators in the quantum Hall theory
\cite{Fradkin:1991nr}
\cite{Cappelli:1993yv}   will be
discussed at the end of this section.

 We are now ready to calculate the corresponding
 Moyal functional form of general non-radially symmetric solitons
 in eq.(15). To be more specific, we examine the unitary transformation,
$U = \exp  [i \hat
 F]$, on the noncommutative scalar function
  \begin{equation}
     \delta \hat \phi = U \hat \phi U^\dagger - \hat \phi \approx i [
\hat
F , \hat \phi ],
  \end{equation}
  where $\hat F$ is the generating function of the $U(\infty)$
transformation, and can be
  expressed in terms of the occupation number basis
   \begin{equation}
     \hat F = \sum_{r,s = 0}^{\infty} \xi_{rs} {\cal J}_{rs},
     \hspace{1cm} {\cal J}_{rs} \equiv ({c^\dagger})^r c^s.
   \end{equation}
   For a Hermitian generator, $\hat F = {\hat F}^\dagger$, the group
parameters
  satisfy $\xi_{rs}^* = \xi_{sr}$. One can show that
  \begin{eqnarray}
   (-i) \delta_{rs} \hat{\phi}_{mn} \equiv [{\cal J}_{rs} ,
\hat{\phi}_{mn}]
   &=& (c^\dagger)^r c^s |m> <n| - |m> <n|(c^\dagger)^r c^s
       \nonumber \\
   &=& \theta(m-s) \frac{\sqrt{m! (m+r-s)!}}{(m-s)!} \hat{\phi}_{m+r-s,n}
       \nonumber \\
   &-& \theta(n-r) \frac{\sqrt{n! (n-r+s)!}}{(n-r)!}
   \hat{\phi}_{m,n-r+s},
  \end{eqnarray}
  where $\theta(n) = 1$ if $n \geq 0$, and $\theta(n) = 0$ otherwise.

  By using eqs.(21),(26), one can calculate the corresponding
transformation on the function
  basis of the Moyal space
  \begin{eqnarray}
    (-i) \delta_{rs} {\varphi}_{mn}(z, \bar z)
   &=& \sqrt{2 \pi \theta} \
  [ \ \theta(m-s) \frac{\sqrt{m!}}{(m-s)! \sqrt{n!}} (a^\dagger)^{m+r-s}
(b^\dagger)^{n}
    \nonumber \\
   &-& \ \theta(n-r) \frac{\sqrt{n!}}{(n-r)! \sqrt{m!}} (a^\dagger)^m
(b^\dagger)^{n-r+s}]
     \Phi_{00} (z, \bar z) \nonumber \\
   &=& [ (a^\dagger)^r a^s - (b^\dagger)^s b^r ] \varphi_{mn}(z, \bar z).
  \end{eqnarray}
  One can thus easily identify the corresponding symmetry generators in
the Moyal
  space to be
  \begin{equation}
    {\cal J}_{rs} \ \  \leftrightarrow \ \ {\cal L}^{a}_{rs} - {\cal
L}^{b}_{sr}
   \equiv  (a^\dagger)^r a^s - (b^\dagger)^s b^r.
  \end{equation}
   Similar consideration for complex scalar function in eq (16)
   gives us
  \begin{equation}
    {\cal J}_{rs} \ \  \leftrightarrow \ \  {\cal L}^{b}_{rs}
   \equiv  (b^\dagger)^r b^s.
  \end{equation}
If we have used the left-handed gauge transformation in eq.(16), the
corresponding generators will be $(a^\dagger)^r a^s$.

  It is interesting to note that either ${\cal L}^{b}$ or ${\cal L}^{a}$
in eq.(28) generates a $W_{1+\infty}$ algebra
\cite{Cappelli:1993yv} \cite{Cappelli:1993ea}
\cite{Cappelli:1993kf} \cite{Iso:1992aa} \cite{Karabali:1994gn}
\cite{Karabali:1994sg}, respectively,

\begin{equation}
        [ {\cal L}_{mn}, {\cal L}_{kl} ] = \sum_{s}^{\min(n,k)}
        \frac{(n+1)!(k+1)!}{(n-s)! (s + 1)! (k-s)!}
        {\cal L}_{m+k-s,n+l-s}
        -(m \leftrightarrow k, n \leftrightarrow l).
      \label{eq:walgebra}
\end{equation}
 This infinite symmetry was discovered some time ago in the study of
quantum Hall theory, where the underlying physics is described by
the Landau problem \cite{Fradkin:1991nr}.

  The Hamiltonian of the Landau problem in the symmetry gauge is
\begin{equation}
  H = - \frac{{\hbar}^2}{2 \mu} ( \vec p - e \vec A)^2 = h \omega
(a^\dagger a +1/2),
  \hspace{0.7cm}
 \vec A = ( - \frac{1}{2} B y, \  \frac{1}{2} B x),
\end{equation}
where the cyclotron frequency is $\omega = \frac{e B}{\mu}$, and the
noncommutative parameter
$\theta$ is related to the magnetic field by
\begin{equation}
 \theta = \frac{4 \hbar}{e B}.
\end{equation}

In eq.(31), $a$ and $a^\dagger$ as defined in eq.(22) are the mechanical
momenta of the electron, and $W_{1+\infty}^a$ can be interpreted
as the energy generating algebra. On the other hand, the physical
meaning of $b$ and $b^\dagger$ defined in eq.(22) are the
so-called magnetic translations of the Hall system. They both
commute with the Hamiltonian due to the trivial translational
symmetry of the Hall system. The projective representation of
these two noncommutative translations is given by
\cite{Fradkin:1991nr}
\begin{equation}
        T_{\xi, \bar \xi} \equiv \exp[\xi b^\dagger - \bar \xi b]
        \hspace{0.5cm} \Rightarrow \hspace{0.5cm}
        T_{\xi, \bar \xi} \ T_{\eta, \bar \eta} =
        \exp[\frac{\xi \bar \eta - \bar \xi \eta}{2}] \
        T_{\xi + \eta, \bar \xi + \bar \eta}.
\end{equation}
It is remarkable that one can extend this symmetry to an infinite
symmetry due to the noncommutativity of these two magnetic
translations. Defining an infinite number of conserved charges as
in eq.(29), it is not difficult to see that they form the
$W_{1+\infty}$ algebra in eq.(30).

 One geometric implication of these two $W_{1+\infty}$
algebras is the manifestation of the well-known area preserving
property of each Landau states as will be discussed in the next
section. We thus have shown that the star-product representation
of $U(\infty)$ symmetry of eq.(15) or eq.(16) in the
non-commutative field theory is generated by two
  $W_{1+\infty}$ algebras, $W_{1+\infty}^a$ and $W_{1+\infty}^b$,
  of the Landau problem. One obvious application is to use this
  algebra to explicitly construct the deformed GMS solitons.

There exists a well-known solution generating technique
\cite{Harvey:2000jb} to generate new solitons with different
topological charges. In particular, instead of using the unitary
$U$ in eq.(24), one introduces the non-unitary isometry or shift
operator,
\begin{equation}
 S_k \equiv \sum_{p=0}^{\infty} |p+k> <p|,
\end{equation}
and the new solitons generated in the Moyal-space will be
\begin{equation}
 S_k {\hat \phi}_{mn} S_k^\dagger \Rightarrow
 \sqrt{\frac{2 \pi \theta \ m! \ n!}{(m+k)!(n+k)!}}
({a^\dagger})^k ({b^\dagger})^k \Phi_{mn}.
\end{equation}

\section{Geometrical Meaning of the $U(\infty)$ Symmetry in
Noncommutative Field Theory}

  Having established the connections between $U(\infty)$ symmetry in the
noncommutative field
theory and $W_{1+\infty}$ algebras in the quantum Hall theory, it
is instructive to illustrate the concrete geometrical meaning of
this infinite symmetry.
  For this purpose, it is sufficient to examine the finite
subgroup transformation of the $U(\infty)$ symmetry on the
noncommutative solitons in eq.(14).
  This finite group is defined by a truncation of the generating function
of the symmetry
\begin{equation}
   U = exp[ i \hat F (c, c^\dagger) ], \hspace{1cm} {\hat F}^\dagger =
\hat F,
\end{equation}
 up to second order in $c$ and $c^\dagger$,
\begin{equation}
     \hat F (c, c^\dagger) = \alpha_0
                          + \xi c^\dagger + \bar \xi c
                          + \beta_1 (c^\dagger c) + \beta_2 (c^\dagger -
c)^2 + \beta_3
    [(c^\dagger)^2 + c^2 ].
\end{equation}
  Due to the simple commutation relation, $[c, c^\dagger]=1$,
  one can verify that this finite set of generators is closed under
 group multiplication. We shall evaluate the induced
 transformations of this subgroup on the noncommutative scalar
 function.
  To solve for the induced transformations, we first decompose any
elements in the subgroup into four independent transformations,

 \vspace{0.5cm}

 $\begin{array}{clll}
 1. & ${\bf translation:}$ &
 U_T(\xi) \equiv \exp [\xi c^\dagger - \bar \xi c],  &
 U_T \ \vec{\bf x} \ U_T^\dagger = \vec {\bf x} - \vec \xi.
 \vspace{0.2cm} \\
 2. & ${\bf rotation:}$ &
 U_R(\theta) \equiv \exp [ i \theta c^\dagger c],  &
 U_R \ \vec{\bf x} \ U_R^\dagger = R \vec {\bf x}. \vspace{0.2cm} \\
 3. & ${\bf squeeze:}$ &
 U_S(\alpha) \equiv \exp [\frac{\alpha}{2} (c^2 -(c^\dagger)^2)], &
 U_S \ \vec{\bf x} \ U_S^\dagger = (e^{\alpha} \hat x, e^{-\alpha}\hat
y).
 \vspace{0.2cm} \\
 4. & ${\bf shearing:}$ &
 U_H(\beta) \equiv \exp [ - i \frac{\beta}{4} (c^\dagger - c)^2], &
 U_H \ \vec {\bf x} \ U_H^\dagger = (\hat x + \beta \hat y, \hat y).
 \end{array}$ \vspace{0.5cm}

 It is easy to check that
 \begin{equation}
  U e^{i \vec k \vec {\bf x}} U^{\dagger} =
  e^{i \vec k \cdot (V \vec {\bf x}) },
 \end{equation}
 where $V$ represents the corresponding transformations,
translation,rotation..etc on the
plane. The induced transformations of the scalar function is then
 calculated to be

\begin{equation}
 \hat{\phi}_U (\vec {\bf x}) \equiv
 U \hat{\phi}(\vec {\bf x}) U^\dagger
 \equiv \int \frac{d^2 \vec k}{(2 \pi)^2} e^{i \vec k \vec {\bf x}}
     \tilde \varphi_U (\vec k) \hspace{0.5cm} \Rightarrow \hspace{0.5cm}
    \varphi_U (\vec x)
   = \varphi (V \vec x).
 \end{equation}

  The induced transformations above generate a finite subgroup of ${\it
w}_{1+\infty}$, namely,
the two dimensional area-preserving diffeomorphism (APD).
This finite subgroup is defined as the direct product of
 translation group (parameterized by a vector
  $\vec \xi$) and the $SL(2,R)$ group (parameterized by a $2 \times 2$
matrix $M$ with unit
  determinant), and the action of group element on the two-dimensional
coordinates is given by
    \begin{equation}
     V(x,y) = M \vec x + \vec \xi = (a x + b y + \xi_x, c x + d y +
\xi_y),
     \hspace{0.8cm} \det M = a d - b c = 1.
    \end{equation}
     The subgroup has an identity element $(M=1, \vec \xi = 0)$, and one
can
    show that the composition rule defines a closed algebra under group
multiplication.
   \begin{equation}
    V_1 \circ V_2 \equiv (M_2, {\vec \xi}_2) \circ (M_1, {\vec
    \xi}_1) = (M_2 \cdot M_1, \ M_2 {\vec \xi}_1 + {\vec \xi}_2).
   \end{equation}
   Finally, the associativity follows naturally from the definition of
group
   multiplication.

 Since there are only three independent parameters in $SL(2,R)$, we can
decompose any group
 element in the $SL(2,R)$
by three independent
 transformations,
 \begin{equation}
   M = M_R \cdot M_H \cdot M_S,
 \end{equation}
where
\begin{equation}
 M_R \equiv \left( \begin{array}{rl} \cos \theta & \sin \theta \\
                                     - \sin \theta & \cos \theta
                       \end{array}     \right), \hspace{0.5cm}
 M_H \equiv \left( \begin{array}{cc} 1 & \beta \\
                                       0 &  1
                       \end{array}     \right), \hspace{0.5cm}
 M_S \equiv \left( \begin{array}{cc}  e^{\alpha} & 0 \\
                                               0   & e^{- \alpha}
                       \end{array}     \right).
\end{equation}
These three transformations represent rotation, shearing, and
squeeze, respectively. Given any $SL(2,R)$ matrix, one can solve
 the three independent
 group parameters as follows,
 \begin{equation}
    M = \left( \begin{array}{cc} m_{11} & m_{12} \\
                                 m_{21} & m_{22}
                 \end{array}     \right) \hspace{0.5cm}
     \rightarrow \hspace{0.5cm} \left\{
     \begin{array}{l}
      \alpha = \frac{1}{2} \ln (m_{11}^2 + m_{21}^2) \vspace{0.2cm} \\
      \beta  = m_{11} m_{12} + m_{21} m_{22}  \vspace{0.2cm} \\
      \theta = \arctan \left( - \frac{m_{21}}{m_{11}} \right)
     \end{array}. \right.
 \end{equation}

 Finally, we list the generating functions associated with these
transformations:
 \begin{enumerate}
   \item {\bf translation}
        \begin{equation}
         \delta x = \xi_x =   \frac{\partial F_T}{\partial y},
             \hspace{0.5cm}
         \delta y = \xi_y = - \frac{\partial F_T}{\partial x},
             \hspace{0.5cm}
         \Rightarrow \hspace{0.2cm} F_T = - \xi_y x + \xi_x y.
        \end{equation}
   \item {\bf rotation}
         \begin{equation}
         \delta x =   \theta y =   \frac{\partial F_R}{\partial y},
             \hspace{0.5cm}
         \delta y = - \theta x = - \frac{\partial F_R}{\partial x},
             \hspace{0.5cm}
         \Rightarrow \hspace{0.2cm} F_R = \frac{\theta}{2} (x^2 + y^2).
         \end{equation}
   \item {\bf squeeze}
         \begin{equation}
         \hspace{-1.2cm}
         \delta x =   \alpha x =   \frac{\partial F_S}{\partial y},
              \hspace{0.5cm}
         \delta y = - \alpha y = - \frac{\partial F_S}{\partial x},
              \hspace{0.5cm}
         \Rightarrow \hspace{0.2cm} F_S = \alpha x y.
         \end{equation}
   \item {\bf shearing}
         \begin{equation}
         \hspace{-1.1cm}
          \delta x = \beta y =   \frac{\partial F_H}{\partial y},
              \hspace{0.7cm}
          \delta y = 0 = - \frac{\partial F_H}{\partial x},
              \hspace{1.1cm}
          \Rightarrow \hspace{0.2cm} F_H = \frac{\beta}{2} y^2.
         \end{equation}
 \end{enumerate}
 One can now use these transformations to explicitly generate new
 solitons. For example, we can use squeeze and shearing to generate
 nonradially symmetric solitons. One other interesting case is
 the boosted solitons \cite{Bak:2000ym}, which has been discussed in the
literature.
 It has been known that Lorentz boost accompanied by a rescaling
 of $\theta$ remains a symmetry of noncommutative field theory. The
 boosted soliton is easily calculated to be
  \begin{equation}
  \phi_n^v (x, y, t; \theta) = \phi_n ( \gamma (x- v t), \ y; \gamma
  \theta )= \phi_n ( \sqrt{\gamma} (x- v t), \  \frac{y}{\sqrt
  \gamma}; \ \theta)
  \end{equation}
 where $\gamma$ is the Lorentz boost parameter and $\phi_n$ are the
radially symmetrical GMS solitons, eq.(14). The first equality of
eq.(49) defines the Lorentz boost of noncommutative field theory.
It is now clear from the second equality that
 the boosted soliton in fact results from a combined transformations of
 translation and squeeze (without $\theta$ rescaling)  discussed in this
section. The key
 mechanism that is functioning behind these transformations is the
 principle of area-preserving, or $W_{1+\infty}$ algebra, discussed in
this
 paper. It is worth noting that the boosted solution of {\it finite
  $\theta$} scalar field theory considered in \cite{Bak:2000ym} is
 area-preserving only approximately since the $U(\infty)$ symmetry is
violated
 by the kinetic term.

   \section{Summary and Discussion}

    In this paper, we have identified the $U(\infty)$ symmetries of
noncommutative field
  theory and $W_{1+\infty}$ algebras of quantum Hall theory.
    Many important properties of the $W_{1+\infty}$ algebra, such
  as area-preserving, Moyal noncommutativity and magnetic
  translations etc. can then help us to understand the dynamics
  of D-branes and noncommutative field theory.
    In particular, we have applied this result to explicitly
    construct various deformed noncommutative solitons including
    boosted solitons.
    The result presented here applies to both pure scalar theory with 
 infinite $\theta$ parameter and
  scalar-gauge theories with finite or infinite $\theta$
  parameter. It also justifies the close relation between quantum
  Hall theory and noncommutative field theory as have been
  suggested recently \cite{Susskind:2001fb}.
    Many interplays between these two fast developing fields remain
  to be uncovered.

\begin{flushleft}
 \Large \bf Acknowledgements
\end{flushleft}

    We thank Pei-Ming Ho, Miao Li, Hsieng-Chung Kao and Hyun-Seok Yang
for inspiration, comments
  and discussions.
    This research is supported by National Science Council of Taiwan
under the budget number
  NSC 89-2112-M-009-045.
    The support for string theory group from NCTS and CTS at NTU is also
acknowledged.

\end{document}